\documentclass{article}

\usepackage{arxiv}

\usepackage[utf8]{inputenc} % allow utf-8 input
\usepackage[T1]{fontenc}    % use 8-bit T1 fonts
\usepackage{hyperref}       % hyperlinks
\usepackage{url}            % simple URL typesetting
\usepackage{booktabs}       % professional-quality tables
\usepackage{amsfonts}       % blackboard math symbols
\usepackage{nicefrac}       % compact symbols for 1/2, etc.
\usepackage{microtype}      % microtypography
\usepackage{lipsum}
\usepackage{graphicx}
\usepackage{todonotes}
\usepackage{paralist}
\usepackage{multirow}
\graphicspath{ {./images/} }

\title{Towards Evolution Capabilities in Data Pipelines}

\author{
 Kevin Kramer \\
  Chair of Databases and Information Systems\\
  University of Hagen\\
  Germany \\
  \texttt{kevin.kramer@fernuni-hagen.de} \\
}

\begin{document}
\maketitle
\begin{abstract}
Evolutionary change over time in the context of data pipelines is certain, especially with regard to the structure and semantics of data as well as to the pipeline operators. Dealing with these changes, i.e. providing long-term maintenance, is costly. The present work explores the need for evolution capabilities within pipeline frameworks. In this context dealing with evolution is defined as a two-step process consisting of self-awareness and self-adaption. Furthermore, a conceptual requirements model is provided, which encompasses criteria for self-awareness and self-adaption as well as covering the dimensions data, operator, pipeline and environment. A lack of said capabilities in existing frameworks exposes a major gap. Filling this gap will be a significant contribution for practitioners and scientists alike. The present work envisions and lays the foundation for a framework which can handle evolutionary change.
\end{abstract}

% keywords can be removed

\section{Introduction}
The last decade was characterized by ever increasing amounts of data. This also led to new technical demands in the context of data storage, transfer and analysis. In order to cope with these demands complex new systems emerged, which in turn require maintenance. Providing this maintenance is costly and even though the systems themselves might run as expected, changes over time, e.g. to the structure and semantics of data, inevitably induce a need to adjust the systems configuration to restore functionality. One estimate suggests that 50-70\% of the total cost of a long running software system can be attributed to maintenance \cite{maintenance}.
Data pipelines are an intuitive way to structure end-to-end data processing. The corresponding tools and frameworks are used in a wide field of domains and for an extensive amount of diverse applications. Still, they also need costly maintenance whenever change, i.e. evolution happens. Adding evolution capabilities to data pipelines and thereby reducing maintenance cost and human involvement could be a big contribution for scientists and practitioners alike. The current work takes the first step in this direction by collecting requirements needed for such a system and by envisioning a data pipeline framework which fulfills these requirements.

The following sections are structured as follows. Section \ref{sec:evo_pipelines} describes the general concepts and challenges of evolution in data pipelines. Important terminology is defined and related work is shown in this section as well. In Section \ref{sec:pipeline_framework} a pipeline framework with evolution capabilities is envisioned and discussed. A conceptual requirements model, which focuses on these evolution capabilities, is presented in Section \ref{sec:req_model}. Finally, the last section concludes the paper and outlines a roadmap for the community towards a pipeline framework with evolution capabilities.

\section{Evolution in Data Pipelines}
\label{sec:evo_pipelines}
This section provides the basis for the current work by defining important concepts as well as presenting related work. Firstly, data pipelines and their components are introduced. Secondly, data pipeline frameworks including their benefits are showcased. Finally, evolution in the context of data pipelines is defined.
\subsection{Data Pipelines}
\begin{figure*}
  \centering
  \includegraphics[width=0.7\linewidth]{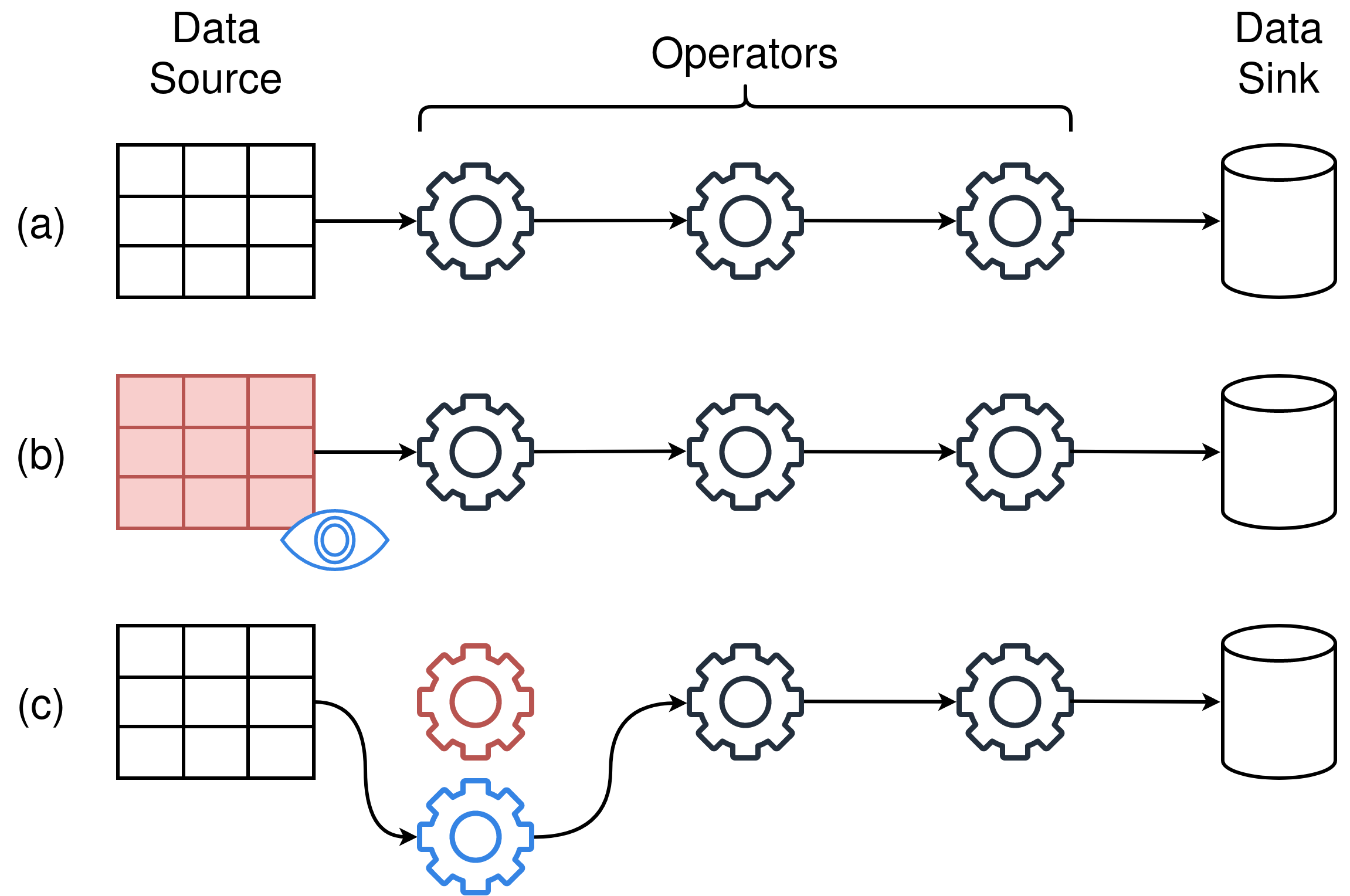}
  \caption{(a) A basic data processing pipeline consisting of a data source, operators and a data sink. (b) Self-awareness: the system perceives a disruption at the data source level. This could be the structural or semantic change of incoming data. (c) Self-adaption: the system automatically adapts to the perceived disruption by swapping the first operator for a different one.}
  \label{fig:pipelines}
\end{figure*}

Data pipelines are used for a plethora of applications and domains such as bioinformatics \cite{DBLP:journals/dase/FjukstadB17, DBLP:journals/bioinformatics/NovellaKHWCBKS19}, manufacturing \cite{DBLP:journals/jbd/IsmailTK19} and cybersecurity \cite{DBLP:journals/cybersec/KoushkiARZGH22}. Broadly speaking, a data pipeline consists of three components: data source(s), operator(s) and data sink(s). Figure~\ref{fig:pipelines} (a) shows such a basic pipeline. 

\cite{DBLP:conf/icse/BiswasWR22} empirically studied the components and stages of 71 \textit{data science} (DS) pipelines \cite{DBLP:conf/icse/BiswasWR22}. Their findings suggest that DS pipelines consist of a \textit{pre-processing} phase, a \textit{model building} phase and a \textit{post-processing} phase. They further extracted tasks and sub-tasks associated with these phases. Subtasks are atomic operators in the context of a pipeline. The pre-processing phase consists of the tasks \textit{data acquisition}, \textit{data preparation} and \textit{storage} which represent the typical components of data engineering and also includes the data source(s). The model building phase is comprised of the tasks \textit{feature engineering}, \textit{modeling}, \textit{training}, \textit{evaluation} as well as \textit{prediction}. These tasks correspond to basic \textit{machine learning} (ML) and \textit{data mining} (DM) functions. The tasks included in the post-processing layer are \textit{interpretation}, \textit{communication} and \textit{deployment} as well as all data sinks. The empirical results show that the pre-processing and the model building phases appeared in 96\% of examined DS pipelines, the post-processing phase only appeared in 52\% of pipelines.

Pipelines can be linear, i.e. one data source, a chain of operators and finally one data sink. \cite{DBLP:journals/sigmod/PsallidasZKHIKK22} empirically studied 8M Jupyter notebooks\footnote{https://www.jupyter.org/} from GitHub\footnote{https://www.github.com/} \cite{DBLP:journals/sigmod/PsallidasZKHIKK22}. Their results which were produced by mining and analyzing the \textit{abstract syntax trees} of all notebooks suggest that 80\% of the pipelines are linear. The structure of pipelines can be interpreted as a \textit{directed acyclic graph} (DAG), allowing for pipelines, which can include several data sources and sinks as well as branching operators, i.e. operators which have more than one input or output. A widespread example of such non-linear data processing are \textit{extract transform load} (ETL). They are used to extract data from multiple heterogeneous sources, transform them to use a common schema and then load them into a data sink such as a data warehouse (which may become a data source itself in the following steps) \cite{DBLP:books/daglib/p/Vassiliadis11}. Even though pipelines can be created using only functions and modules by chaining their inputs and outputs together \cite{DBLP:journals/sigmod/PsallidasZKHIKK22}, pipeline frameworks allow users to generate, maintain and administrate complex pipelines.

\subsection{Pipeline Frameworks}
The number of existing pipeline frameworks is overwhelming. A popular collection of pipeline tools at GitHub\footnote{https://www.github.com/pditommaso/awesome-pipeline} includes 122 pipeline frameworks. At the same time there is almost no scientific attention on the abstract concepts of these systems. Some conceptual work was made by \cite{DBLP:journals/corr/abs-1901-01908} \cite{DBLP:journals/corr/abs-1901-01908}. The author proposes an important distinction in order to categorize pipeline frameworks. He divides frameworks into \textit{task-driven} and \textit{data-driven}. Task-driven frameworks are agnostic about actual data and operations that occur during a pipeline run. Their focus lies on managing inter- and intra-pipeline dependencies and scheduling large numbers of pipelines in parallel. Popular proponents of this category are Luigi\footnote{https://www.github.com/spotify/luigi} and Apache Airflow\footnote{https://www.airflow.apache.org/}. Data-driven pipelines are -- to a varying degree -- aware of the data they process and the included operations. These frameworks put a focus on data (and operator) lineage also called  provenance, i.e. they allow the user to retrace the history of a data artifact by saving and curating metadata on all steps of the artifact producing pipeline. A popular data-driven framework which logs  various metadata during pipeline runs is Dagster\footnote{https://www.dagster.io/}. Some frameworks in this category enable data provenance by using a version control system similar to Git\footnote{https://www.git-scm.com/}. A prominent example of this is Pachyderm\footnote{https://www.pachyderm.com/}.

Comparing pipeline frameworks is made difficult by a number of factors: the sheer amount of different frameworks, the lack of a theoretical basis for analysis, the overlapping functionality and the differing ways to achieve the same goal within two frameworks. A thorough search of related work and literature focusing on such comparison, only revealed one paper \cite{DBLP:conf/rcdl/MatskinTLPTNR21}. Even though the analysis was geared towards a specific system and its requirements, the general results and especially the comparison criteria are a helpful first step towards distinguishing pipeline frameworks. Some of these criteria and their possible values include:
\smallskip
\begin{compactitem}
    \item \textbf{Type:} business, science, big data
    \item \textbf{Model:} script-based, event-based, adaptive, declarative and procedural
    \item \textbf{Separation of concerns:} asks whether or not high-level pipeline definitions can be separated from low-level data and operator implementations
    \item \textbf{Language:} \textit{general purpose language} (GPL), \textit{domain specific language} (DSL)
    \item \textbf{Pipeline programming:} text-based, graphical, visual
    \item \textbf{Reusability:} asks whether or not a framework provides tools for reusing existing pipeline definitions as well as individual components of a previously defined pipeline
    \item \textbf{Containerization:} asks if pipeline components, whole pipelines and the pipeline framework itself can be deployed in a container
    \item \textbf{Monitoring:} asks whether or not the framework allows for runtime observation of the system or if it is granting logging capabilities
\end{compactitem}
\smallskip
Some of these results are referenced in Section \ref{sec:pipeline_framework}. In Section \ref{sec:req_model} these basic criteria are extended with a special focus on evolution capabilities. The particularities resulting from evolution will be presented in more detail in the next subsection.

\subsection{Pipeline Evolution}
\label{subsec:pip_evo}
Evolution means change over time. In the realm of computer science change can mean a lot of different things. The emergence and widespread adoption of a new data format (such as \textit{JSON}\footnote{https://www.json.org}) or programming model (such as \textit{MapReduce} \cite{DBLP:conf/osdi/DeanG04}) are examples of this. This type of evolution is often gradual and influenced by many different factors. In the context of data pipelines and corresponding frameworks evolution can happen over different time frames, ranging from gradual to sudden. The main evolution factors are so-called \textit{disruptors}, which can affect all components and their interactions with each other. The changes triggered by disruptors are diverse, but can be broadly categorized into data, operator and environment disruptors. 

The structure and semantics of data might change, affecting data sources and sinks as well as data artifacts created within the pipeline, e.g. interim results. Structural changes in data might occur over time due to altered data producers or operators. Semantic changes in data can emerge from technical, legislative but also societal reasons.

Operator functionality might also experience evolution, e.g after a software update, resulting in different APIs or a changed set of available (hyper)parameters. Another form of change in this context is choosing a different operator for a specific task which accepts the same input as the old one but produces a different output, e.g. a different data structure. This leads to the need to adapt the pipeline to fit this new operator.

Also, the environment in which the pipeline is run can change over time. For example, the hardware could change resulting in more processing power or more cluster nodes becoming available. Adapting to such change by increasing the number of pipelines running in parallel or utilizing bigger batch sizes in order to increase efficiency could be possible examples.

\section{Pipeline Framework with Evolution Capabilities}
\label{sec:pipeline_framework}
\begin{figure*}
  \centering
  \includegraphics[width=0.7\linewidth]{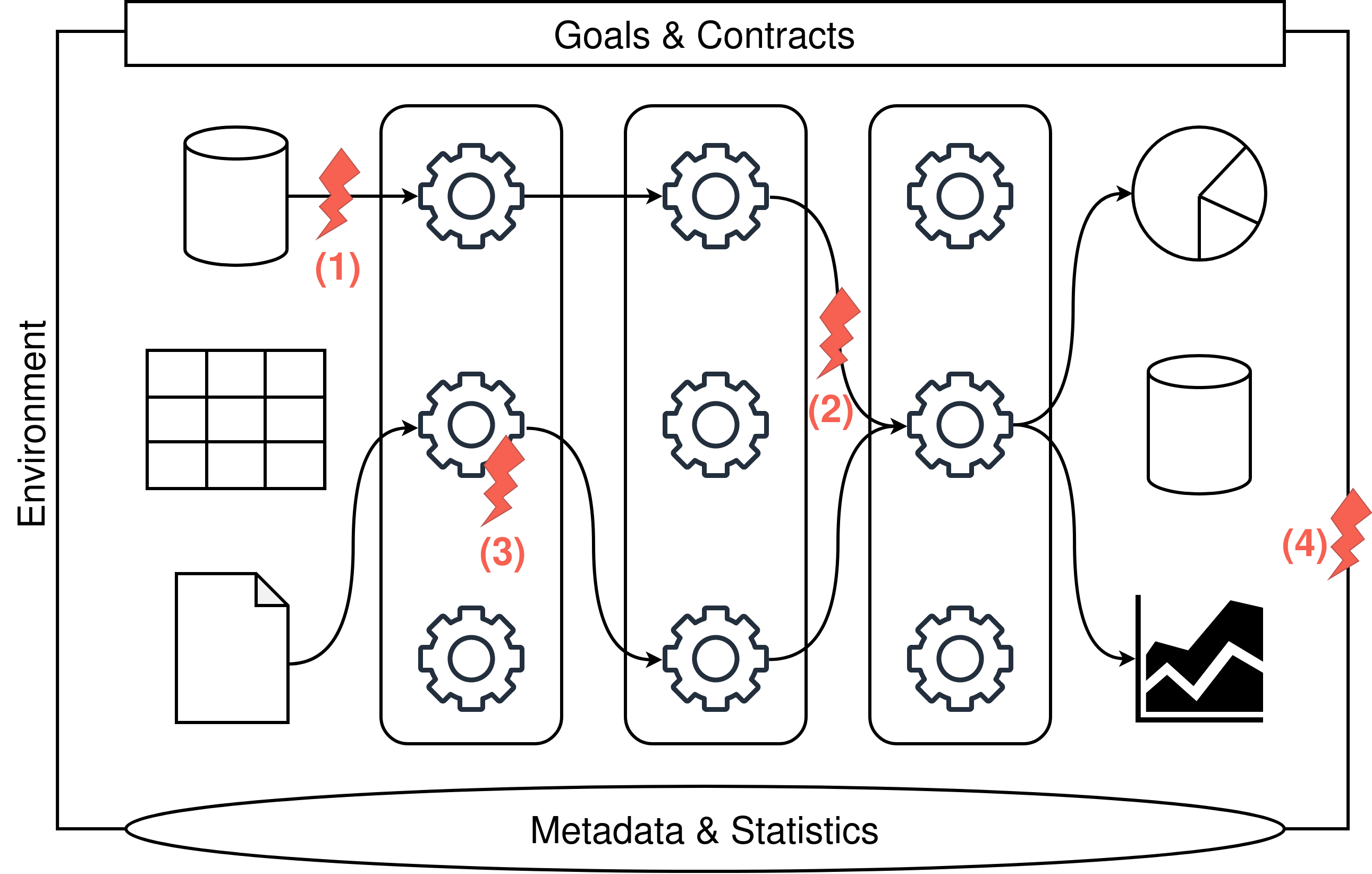}
  \caption{Pipeline framework and its components. Evolution can happen in the form of structural and semantic changes to the data during loading (1) and through operator processing (2) as well as to operators (3), e.g. after a software update. The environment, i.e. hardware, scaling, etc., might also change over time (4). Based on a figure from \cite{DBLP:journals/dbsk/KlettkeS22}.}
  \label{fig:framework}
\end{figure*}

In this section a pipeline framework with evolution capabilities is envisioned and discussed. Figure~\ref{fig:framework}, based on a figure from \cite{DBLP:journals/dbsk/KlettkeS22}, shows a graphical representation of the proposed framework. The outside of the figure is made up of the environment frame including goals and contracts as well as metadata and statistics. These elements represent the available resources, user objectives and metadata, which the system gathered, stored and aggregated throughout its lifecycle. Within this frame there are essentially five columns. They represent (from left to right) data sources, operators and data sinks. The arrows connect the individual components and show two pipelines, each consisting of a data source, three operators and a data sink. Evolutionary change can happen at several points during a pipeline's lifecycle. In Figure~\ref{fig:framework} these disruption points are shown as red flashes. Structure and semantics of data might change at the data sources as well as within the pipeline. Evolution can also affect the operators and the environment in which the pipelines are run. In any case, an ideal pipeline framework could automatically adapt to these changes. 

Concerning  adaptability, an important distinction needs to be made. Generally speaking, it is possible to build pipelines in existing frameworks, that are very flexible. One class of systems, which are very flexible are \textit{adaptive workflows}, first presented by \cite{DBLP:conf/iceis/AalstBVVV99}. Besides being mainly task-driven, these systems adapt themselves based on strict, predefined rules. An example of such a system is \textit{AdaptFlow} presented in \cite{DBLP:conf/cgp/GreinerRHLMR04}. Given a treatment plan in the medical context, AdaptFlow can notice logical errors and choose a different path in the predefined workflow. This flexibility is completely dependent on and bounded by the treatment workflow. Generally speaking, the space of possible alterations, given such a flexible system, is significantly smaller than the space envisioned in the present work. This stems from the fact that a pipeline framework with evolution capabilities dynamically creates and alters this search space, in order to find an optimal solution, at different times during the system's lifecycle. This demonstrates that flexibility is not the same as adaptability. It is also possible to build \textit{meta pipelines} especially for monitoring changes as well as adapting to these changes. Even though this is currently the most practical solution for achieving evolution capabilities in existing frameworks, this approach does not represent \textit{real} evolution capabilities as they were defined in the previous sections. In any case, before adapting to evolution, the underlying changes need to be noticed and recognized. 

\subsection{Self-awareness}
The first step in dealing with evolution is to be aware of change. Figure~\ref{fig:pipelines} (b) shows this step in dealing with evolution. Data-driven frameworks are usually more aware of change than task-driven ones since they provide more monitoring capabilities and allow for concepts such as reproducibility and provenance which are closely related to evolution. A tool for inspecting pipelines which runs on existing Python code is \textit{mlinspect} \cite{DBLP:conf/cidr/GrafbergerSS21, DBLP:conf/sigmod/GrafbergerGSS21}. It extracts the DAG structure of a pipeline and helps the user to identify problems and bugs. For example it can help to identify a skewed data distribution which would lead to unfair \cite{DBLP:journals/pvldb/AsudehJ20a} results. \textit{ArgusEyes} \cite{DBLP:conf/cidr/SchelterGGSK022} is a tool for inspecting classification pipelines which builds upon mlinspect. It enables the user to check whether best practices are applied while also providing various metadata to analyze pipelines. Even though these tools are not intended to track the evolution of pipelines and their components, but rather focus on helping practitioners with a specific issue, the underlying architecture can serve as useful guidance for the development of a pipeline framework with evolution capabilities. Another important aspect is to track data changes across pipeline steps. The authors of \cite{DBLP:journals/dbsk/KlettkeLS21} present three measuring approaches that are utilized in order to deal with bias. 

Monitoring capabilities, gathering and storing metadata as well as calculating and providing statistics on these findings are critical functionalities towards evolution capabilities in pipeline frameworks. They are necessary in all dimensions and are the basis for self-awareness. Tools like mlinspect and ArgusEyes, but also existing data-driven frameworks like Dagster can be a starting point towards achieving such functionality. Perceiving change in operator results or contracts leading to the automatic swapping or parameter change is also fundamentally important. One project that can be of help in this regard is IBM Lale \cite{baudart_et_al_2021} which automatically creates optimal pipelines based on scikit-learn\footnote{https://www.scikit-learn.org/stable/} functions. Once the system is aware of change, it needs to adapt to the new circumstances.

\subsection{Self-adaption}
Automatic acting upon change can only be done with respect to a goal. This goal could be as simple as ensuring functionality and as complex as automatically optimizing the performance and accuracy of several big data pipelines running in parallel given certain hardware. Figure~\ref{fig:pipelines} (c) shows the adaption step, after a disruption has been perceived by the self-awareness capabilities. In this context it is decisive to formulate a goal including a fitting representation, which the pipeline framework can use to evaluate decisions. The dimensions for pipeline and environment shown in the last section both contain the evolution requirement to provide an interface for goals. This reveals a potential conflict: A pipeline with the goal to achieve the best possible accuracy for a ML task might want to simulate a lot of different pipelines to find the best one and to achieve this goal. At the same time simulations and tests might cost a lot of computational resources, which could stand in contrast to the environment dimension's goal to provide a certain performance to all pipelines. A pipeline framework with evolution capabilities needs to have dynamic functionality to deal with these kinds of conflicts.

The vision of self-adapting systems is not unique to the present work. The authors of \cite{DBLP:journals/dbsk/KlettkeS22} present four generations in data engineering for data science ranging from simple data pre-processing to fully automated data curation. In \cite{DBLP:conf/ideas/HolubovaKL22} the authors envision a framework for multi-model databases, which is self-adapting with regard to design and maintenance. Similar to the insight gained from tools like mlinspect and ArgusEyes in the context of evolution awareness, other self-adaptive systems can help to understand the underlying components and their interplay. For example \cite{DBLP:journals/dpd/HillenbrandSNK22} propose a system which automatically chooses an optimal data migration strategy given some constraints like service-level agreements \cite{DBLP:journals/dpd/HillenbrandSNK22}. Pachyderm which runs natively in Kubernetes\footnote{https://kubernetes.io/} has a built-in system for distributed computing / scaling, which is very simple and should be considered in the context of the environment dimension. The empirical results of \cite{DBLP:conf/rcdl/MatskinTLPTNR21} showed a complete lack of a simulation environment in all studied frameworks. Simulation and the use of synthetic data \cite{DBLP:conf/acit3/AbufaddaM21} are important components, which need to be incorporated especially for the pipeline and environment dimensions since their self-adaption strategies need a search space to optimize towards a goal.

\section{Conceptual Requirements Model}
\label{sec:req_model}
As described in the previous sections, there is no framework with comprehensive evolution capabilities yet. This emphasizes the need for a requirements model, encompassing important components and their interplay as well as system functionalities. The model presented in this section is conceptual, i.e. it was not derived through a structured method from the field of requirements engineering \cite{DBLP:conf/se/0001FFVKWPCCGLM21}. It rather evolved from technical talks with experienced colleagues and a rough analysis and comparison of existing pipeline frameworks. It can serve as the inception step for a structured requirements gathering process and furthermore helps with the testing of existing frameworks for their evolution capabilities.

The requirements are structured into two categories, self-awareness and self-adaption as well as four dimensions.

\begin{compactitem}
    \item \textbf{Data:} Data sources and sinks, structure and semantics of data
    \item \textbf{Operator:} Modules and functions and their respective inputs and outputs
    \item \textbf{Pipeline:} Creation and administration of pipelines
    \item \textbf{Environment:} Available hardware and scheduling, scaling and orchestration of pipelines
\end{compactitem}

Table \ref{tab:reqs} presents an overview of the requirements. The following sections describe the requirements listed in Table \ref{tab:reqs} in detail.

\begin{table*}
  \caption{Conceptual requirements and their corresponding dimensions, categorized into self-awareness and self-adaption}
  \label{tab:reqs}
  \begin{tabular}{lll}
    \toprule
    \textbf{Category}&\textbf{Requirement}&\textbf{Dimension}\\
    \midrule
    \multirow{12}{7em}{Self-awareness}&Collecting and storing metadata&all\\
    &Versioning of metadata&all\\
    &Versioning of component artifacts&all\\
    &Versioning of configuration files&all\\
    &Providing provenance capabilities&all\\
    &Analyzing metadata and creating statistics&all\\
    &Noticing structural changes&data\\
    &Noticing semantic changes&data\\
    &Noticing changes to contracts,APIs and interfaces&operator\\
    &Noticing changes to available computing resources&environment\\
    &Monitoring processing results and performance&operator, pipeline\\
    &Providing an interface for goal definition&operator, pipeline, environment\\
    \midrule
    \multirow{5}{7em}{Self-adaption}&Initiating an adaption, based on the violation of a goal&operator, pipeline, environment\\
    &Automatically swapping operators&operator, pipeline\\
    &Automatically changing pipeline structure and components&pipeline\\
    &Automatically optimizing resource distribution and scheduling&environment\\
    &Providing a simulation space to test potential alteration&pipeline, environment\\
  \bottomrule
\end{tabular}
\end{table*}

\subsection{Self-awareness Requirements}
Self-awareness means being aware of change. This change is always relative with respect to some previous state, i.e. in order to be self-aware, a system needs to store at least one previous state for comparison with the current state. Therefore, collecting and storing metadata over all dimensions is an integral requirement for a self-aware pipeline framework. Even though comparing two system states is sufficient to notice change, in many cases it would be beneficial to have a history of system states. Creating a versioned history of metadata allows for more complex concepts and techniques to be applied, e.g. extracting (meta)data distributions or using window-based anomaly detection to notice change. Versioning of metadata, component artifacts and configuration files would enable the self-aware system to notice different forms of change and distinguish them. For example it could differentiate between an abrupt change to the interface of an operator after a software update and the gradual decrease of data quality, based on the wrong composition of preprocessing operators. Collecting and storing such data is important, but so is managing and curating it, which leads to the need for provenance capabilities over all dimensions. Also providing tools to analyze metadata, for example to aggregate historic data into statistical values, is an important requirement. Aggregated data enables a different perspective of change.

When looking at the data dimension, the two fundamental requirements a pipeline framework with evolution capabilities has to fulfill are noticing changes to the structure of data and noticing changes to the semantics of data. These disruptors almost always trigger an adaption and therefore, being aware and dealing with them, is of utmost importance. The same can be said about the operator dimension. A changing operator interface will most certainly result in an erroneous pipeline. Hence, noticing such change is a critical requirement. Changes to the environment do not necessarily result in non-functioning pipelines, but rather influence the performance. Still, noticing changes to the environment, e.g. available hardware, is important to achieve framework performance goals, such as optimal utilization of available resources. A similar approach needs to be taken for operator and pipeline goals. Processing results and performance of individual operators as well as pipelines need to be monitored, in order to compare these results to predefined goals. Diverse metrics for goal definition can be imagined, ranging from speed and throughput performance to data quality and model accuracy. This leads to framework requiring an interface for goal definition. This interface allows the user to specify objectives with respect to individual operators, pipelines and the whole framework. At the same time, this goal definition is used for comparison with the current as well as historic states of the system, to notice change and possibly initiate an adaption.
\subsection{Self-adaption Requirements}
Once the system is aware of a significant change, it triggers an adaption. Based on the dimension in which the adaption should occur, i.e. operator, pipeline or environment, the prerequisites for all possible adaption operation are checked. This first step towards an adaption is an important requirement for a pipeline framework with evolution capabilities, since it creates a search space for possible adjustments. The operations, which make up these adjustments, represent crucial requirements as well. They include the automatic swapping of an operator, the automatic change of pipeline structure and/or components, as well as the automatic optimization of resource distribution and pipeline scheduling. The search space of all possible operations is transformed into a simulation space, in which possible alterations are tested. This space connects the user's goal definitions with the self-awareness metadata, while at the same time providing simulation and optimization capabilities, in order to find an optimal adaption.

\section{Conclusion and Future Work}
\label{sec:disc_out}
The present work defined and showcased data pipelines and their corresponding frameworks. Evolution in the context of these systems was introduced and a conceptual requirements model was proposed, comprised of all components of such systems, categorized by self-awareness and self-adaption and structured into four dimensions. By envisioning a system which fulfills these requirements, a first step was made towards a framework, which would need less maintenance based on its self-awareness and self-adaption, i.e. evolution capabilities. This type of framework could be a substantial contribution for scientists and practitioners alike.

The paper is concluded with a set of steps that need to be taken by the community towards achieving evolution capabilities in data pipelines. First of all, a proper requirements model using concepts and methods of requirements engineering must be constructed. This must include a structured requirements gathering process comprised of talking to stakeholders, who would benefit from the proposed system, as well as an in-depth analysis of existing concepts and techniques with regard to self-awareness and self-adaption. As a result, this step would produce a system specification encompassing requirements, including non-functional ones, use-cases and a basic software architecture, as well as formal definitions of new terms. In the next step, these results need to be compared to existing frameworks and tools, in order to find working solutions, but also gaps. All dimensions must be thoroughly analyzed and the system specification must be iteratively adjusted. During this phase software engineering and architecture principles, which support evolution capabilities must be derived from existing systems and be incorporated into the specification. The secondary goal of this step is to either find a framework, which provides a good basis for evolution capabilities -- at least with respect to a certain dimension --, or to discover the need to conceptualize and implement the missing components from scratch. In any case, the next step would be the creation of a prototype. As a final step, this prototype must be evaluated and validated, given the system specification.

\section{Acknowledgments}
  The author wants to thank Meike Klettke, Stefanie Scherzinger, and Uta Störl for many prolific discussions as well as helpful suggestions, with regard to evolution capabilities in data pipelines, without which the present work would not have been possible.

\bibliography{template}
\bibliographystyle{alpha}
\end{document}